\documentclass[10pt,twocolumn]{article}
\usepackage{latexsym,graphicx}
\usepackage{latexsym,graphicx}
\newcommand{\be}{\begin{equation}}
\newcommand{\ee}{\end{equation}}

\catcode `\@=11 \catcode `\@=12
\usepackage[left=2.5cm,top=2.75cm,right=2.5cm,bottom=1.75cm]{geometry}
\begin{document}
\begin{center}
\large{\bf{Cosmological constant dominated transit universe from early deceleration to current 
acceleration phase in Bianchi-V space-time}} \\
\vspace{10mm}
\normalsize{Anil Kumar Yadav}\\
\vspace{4mm}
Department of Physics, Anand Engineering\\
College, Keetham, Agra - 282 007, India.\\
\vspace{2mm}
e-mail: abanilyadav@yahoo.co.in

\vspace{2mm}
\end{center} 
\vspace{10mm}
\begin{abstract} 
The paper presents the transition of universe from early decelerating phase to current accelerating phase 
with viscous fluid and time dependent cosmological constant $(\Lambda)$ as source of matter in Bianchi-V space-time. 
To study the transit behaviour of universe, we have assumed the scale factor as increasing function of time which 
generates a time dependent deceleration parameter (DP). The study reveals that the cosmological term does not 
change its fundamental nature for $\xi$ = constant and $\xi=\xi(t)$, where $\xi$ is the coefficient 
of bulk viscosity. The $\Lambda(t)$ is found to be positive and 
is decreasing function of time. The same is observed by recent supernovae observations. 
The physical behaviour of universe has been discussed in detail.\\      
\end{abstract}
\smallskip

Keywords: Cosmological constant, Bianchi-V space-time, deceleration parameter\\

 PACS: 98.80.Es, 98.80-k 
\section{Introduction}
One of the out standing problem in particle physics and cosmology is the cosmological constant 
problem: its theoretical expectation values from quantum field theory exceed observational limits by 120 
orders of magnitude (Padmanabhan 2003). Even if such high energy are suppressed by super-symmetry, the 
electroweak corrections are still 56 orders highers. This problem was further sharpened by recent 
observation of supernova Ia (SN Ia), which reveal the striking discovery that our universe has lately 
been in its accelerated expansion phase (Riess et al. 1998; Perlmutter et al. 1999) cross checks from 
the cosmic background microwave radiation (CMBR) and large scale structure (LSS), all confirm this unexpected 
result (Riess et al. 2004; Astier et al. 2006). Numerous dynamical dark energy models 
have been proposed in the literature, such as quintessence (Ratra and Peebles 1988), 
phantom (Caldwell 2002), k-essence (Armendariz et al. 2000), tachyon (Padmanabhan 2002), 
DGP (Dvali et al. 2000) and chaplygin gas (Kamenshchik et al 2001). 
However, the simplest and most theoretically appealing candidate for dark energy 
is the vacuum energy (or the cosmological constant $\Lambda$) with a constant equation of state 
parameter equal to $-1$. \\

Experimental study of isotropy of the cosmic microwave background radiation (CMBR) and speculation 
about the amount of helium formed at the early stages of the evolution of universe have stimulated 
theoretical interest in anisotropic cosmological models. At the present state of evolution, 
the universe is spherically symmetric and the matter distributed in it is on the whole isotropic 
and homogeneous. But in its early stages of evolution it could not have such a smoothed out picture 
because near the big bang singularity neither the assumption of spherical symmetry nor of isotropy 
can be strictly valid. Anisotropy of the cosmic expansion, which is supposed to be damped out in the course 
of cosmic evolution, is an important quantity. Recent experimental data and critical arguments 
support the existence of an anisotropic phase of the cosmic expansion that approaches an isotropic one. 
Therefore, it make sense to consider models of universe with anisotropic background. Here we 
confine ourselves of models of Bianchi-type V. Bianchi-V space-time has a fundamental role 
in constructing cosmological models suitable for describing the early stages of evolution of 
universe. In literature Collins (1974), Maarteens and Nel (1978), Camci at al. (2001), Pradhan 
and Rai (2004), Yadav (2009), Yadav et al. (2011) and Yadav (2011a) have studied anisotropic 
Bianchi-V models in different physical contexts. Saha (2006) has presented the quadrature form of metric 
functions for Bianchi-V model with perfect fluid and viscous fluid.\\    
 
The investigation of relativistic cosmological models usually has the cosmic fluid 
as perfect fluid. However, these models do not incorporate dissipative mechanism responsible 
for smoothing out initial anisotropies. It is believed that during neutrinos decoupling, the matter 
behaved like a viscous fluid (Klimek 1976) in early stages of evolution. It has been suggested 
in large class of homogeneous but anisotropic universe, the anisotropy dies away rapidly. The most 
important mechanism in reducing the anisotropy is neutrinos viscosity at temperature just above $10^{10}$K. 
It is important to develop a model of dissipative cosmological processes in general, so that one can analyze 
the overall dynamics of dissipation without getting lost in the detail of complex processes. Coley (1990) studied 
Bianchi-V viscous fluid cosmological models for barotropic fluid distribution. Murphy (1973) has 
investigated the role of viscosity in avoiding the initial big bang singularity. Padmanabhan and 
Chitre (1987) have shown that bulk viscosity leads to inflationary like solution. Pradhan et al (2004, 2005) 
investigated viscous fluid cosmological models in Bianchi-V space-time with varying $\Lambda$. Recently 
Singh and Kale (2009, 2010), Yadav et al (2012) have studied bulk viscous cosmological models with variable $G$ 
and $\Lambda$.\\

In this paper, we have studied the transit behaviour of universe with time dependent $\Lambda$ in 
Bianchi-V space-time. To study the transit behaviour of universe, we have assumed 
the scale factor as increasing function of time which generates a 
time dependent DP. The paper is organized as follows. In section 2, the model and generalized law 
for scale factor have been presented. The section 3 deals with field equations. Some particular models have 
been discussed in section 4 and section 5. The last section 6 contains the concluding remarks. 
\section{Model and generalized law for scale factor that yielding time dependent DP}
We consider the space-time metric of spatially homogeneous and 
anisotropic Bianchi-V of the form
\begin{equation}
 \label{eq1}
ds^{2}=-dt^{2}+A^{2}dx^{2}+e^{2\alpha x}\left(B^{2}dy^{2}+C^{2}dz^{2}\right)
\end{equation}
where $A(t)$, $B(t)$ and $C(t)$ are the scale factors in different spatial directions and 
$\alpha$ is a constant.\\
We define the average scale factor $(a)$ of Bianchi-type V model as
\begin{equation}
 \label{eq2}
a=(ABC)^{\frac{1}{3}}
\end{equation}
The spatial volume is given by
\begin{equation}
\label{eq3}
V = a^{3} = ABC 
\end{equation}
Therefore, the generalized mean Hubble's parameter $(H)$ read as 
\begin{equation}
 \label{eq4}
H=\frac{\dot{a}}{a}=\frac{1}{3}\left(H_{1}+H_{2}+H_{3}\right)
\end{equation}
where $H_{1}=\frac{\dot{A}}{A}$, $H_{2}=\frac{\dot{B}}{B}$ and $H_{3}=\frac{\dot{C}}{C}$ are the 
directional Hubble's parameters in the direction of $x$, $y$ and $z$ respectively. An over dot denotes 
differentiation with respect to cosmic time t.\\

Since metric (\ref{eq1}) is completely characterized by average scale factor therefore let 
us consider that the average scale factor is increasing function of time as following
\begin{equation}
\label{eq5}
a=(t^{n}e^{t})^{\frac{1}{m}} 
\end{equation}
where $m > 0$ and $n \geq 0$ are constant.\\
Such type of ansatz for scale factor has already been considered by Yadav (2012) which 
generalized the one proposed by Pradhan and Amirhashchi (2011). The proposed law (\ref{eq5}) yields 
a time dependent DP which describes the transition of universe from early decelerating phase to 
current accelerating phase.\\

The value of DP (q) for model (\ref{eq1})is found to be
\begin{equation}
\label{eq6}
q=-\frac{\ddot{a}a}{\dot{a}^{2}} = -1+\frac{mn}{(n+t)^{2}}
\end{equation}
From equation (\ref{eq6}), it is clear that the DP $(q)$ is time dependent. 
Also, the transition redshift from deceleration expansion to accelerated expansion 
is about $0.5$. Now for a universe which was decelerating in past and accelerating at 
present time, DP must show signature flipping (Amendola 2003; Riess et al. 2001).\\

\section{Field equations}
For the bulk viscous fluid, the energy momentum tensor is given by
\begin{equation}
\label{eq7} 
T^{i}_{j} = (\rho + \bar{p})v^{i}v_{j} - p g^{i}_{j} 
\end{equation}
where $\rho$ is the energy density, $\bar{p}$ is the effective pressure of 
the fluid, and $v^{i}$ is the fluid four velocity vector. In co moving 
system of co-ordinates, we have $v^{i} = (1,0,0,0)$. The effective pressure $\bar{p}$ 
is related to the equilibrium pressure p by (Yadav 2011b)
\begin{equation}
\label{eq8}
\bar{p}=p-3\xi H
\end{equation}
where $\xi$ is the coefficient of bulk viscosity that determines the magnitude of 
viscous stress relative to expansion.\\

The Einstein's field equations with cosmological constant (in gravitational units $c = 1$, $8\pi G = 1$) read as
\begin{equation}
\label{eq9}
R_{j}^i - \frac{1}{2}g_{j}^{i}R  = -T_{j}^i + \Lambda g^{i}_{j}
\end{equation}
The Einstein's field equations (\ref{eq9}) for the line-element (\ref{eq1}) 
lead to the following system of equations 
\begin{equation}
 \label{eq10}
\frac{\ddot{B}}{B}+\frac{\ddot{C}}{C}+\frac{\dot{B}\dot{C}}{BC}-\frac{\alpha^{2}}{A^{2}} = -\bar{p} + \Lambda
\end{equation}
\begin{equation}
 \label{eq11}
\frac{\ddot{A}}{A}+\frac{\ddot{C}}{C}+\frac{\dot{A}\dot{C}}{AC}-\frac{\alpha^{2}}{A^{2}} = -\bar{p}+\Lambda
\end{equation}
\begin{equation}
 \label{eq12}
\frac{\ddot{A}}{A}+\frac{\ddot{B}}{B}+\frac{\dot{A}\dot{B}}{AB}-\frac{\alpha^{2}}{A^{2}} = -\bar{p}+\Lambda
\end{equation}
\begin{equation}
 \label{eq13}
\frac{\dot{A}\dot{B}}{AB}+\frac{\dot{A}\dot{C}}{AC}+\frac{\dot{B}\dot{C}}{BC}-\frac{3\alpha^{2}}{A^{2}} = \rho + \Lambda
\end{equation}
\begin{equation}
 \label{eq14}
\frac{2\dot{A}}{A}-\frac{\dot{B}}{B}-\frac{\dot{C}}{C} = 0
\end{equation}
Combining equations (\ref{eq10})-(\ref{eq13}), one can easily obtain continuity equation as 
\begin{equation}
\label{eq15}
\dot{\rho}+(\rho+\bar{p})\left(\frac{\dot{A}}{A}+\frac{\dot{B}}{B}+\frac{\dot{C}}{C}\right)+\dot{\Lambda} = 0
\end{equation}
Integrating equation (\ref{eq14}) and absorbing the constant of integration into $B$ or $C$, we obtain
\begin{equation}
\label{eq16}
A^{2}=BC
\end{equation}
Subtracting equations (\ref{eq10}) from (\ref{eq11}), (\ref{eq10}) from (\ref{eq12}), (\ref{eq11}) 
from (\ref{eq12}) and taking second integral of each, we obtain the following three relations respectively
\begin{equation}
\label{eq17}
\frac{A}{B} = b_{1}\;exp\left(x_{1}\int a^{-3}dt\right)
\end{equation}
\begin{equation}
\label{eq18}
\frac{A}{C} = b_{2}\;exp\left(x_{2}\int a^{-3}dt\right)
\end{equation} 
\begin{equation}
\label{eq19}
\frac{B}{C} = b_{3}\;exp\left(x_{3}\int a^{-3}dt\right)
\end{equation}
where $b_{1}$, $b_{2}$, $b_{3}$, $x_{1}$, $x_{2}$ and $x_{3}$ are constant of integrations.\\

From equations (\ref{eq16}) - (\ref{eq19}) and (\ref{eq5}), the metric functions can be explicitly 
written as
\begin{equation}
\label{eq20}
A(t) = (t^{n}e^{t})^{\frac{1}{m}}
\end{equation}
\begin{equation}
\label{eq21}
B(t)=d\;(t^{n}e^{t})^{\frac{1}{m}}\;exp\left(\ell\int(t^{n}e^{t})^{-\frac{3}{m}}dt\right)
\end{equation}
\begin{equation}
\label{eq22}
C(t)=d^{-1}\;(t^{n}e^{t})^{\frac{1}{m}}\;exp\left(-\ell\int(t^{n}e^{t})^{-\frac{3}{m}}dt\right)
\end{equation}
where $d = (b_{2}b_{3})^{\frac{1}{3}}$, $\ell = \frac{x_{2}+x_{3}}{3}$ with $b_{2} = b_{1}^{-1}$ and $x_{2} = - x_{1}$.\\

The physical parameters such as scalar of expansion $(\theta)$, spatial volume $(V)$, anisotropy parameter 
$(\bar{A})$, shear scalar $(\sigma^{2})$ and directional Hubble's parameters $(H_{x}, H_{y}, H_{z})$ 
are respectively given by
\begin{equation}
\label{eq23}
\theta = 3H = \frac{3}{m}\left(\frac{n}{t}+1\right)
\end{equation}
\begin{equation}
\label{eq24}
V = (t^{n}e^{t})^{\frac{3}{m}}
\end{equation}
\begin{equation}
\label{eq25}
\bar{A} = \frac{2\ell^{2}m^{2}}{3}(t^{n}e^{t})^{-\frac{6}{m}}\left(\frac{n}{t}+1\right)^{-2}
\end{equation}
\begin{equation}
\label{eq26}
\sigma^{2} = \frac{2\ell^{2}}{3}(t^{n}e^{t})^{-\frac{6}{m}}
\end{equation}
\begin{equation}
\label{eq27}
H_{x} = \frac{1}{m}\left(\frac{n}{t}+1\right)
\end{equation}
\begin{equation}
\label{eq28}
H_{y} = \frac{1}{m}\left(\frac{n}{t}+1\right)+\ell(t^{n}e^{t})^{-\frac{3}{m}}
\end{equation}
\begin{equation}
\label{eq29}
H_{z} = \frac{1}{m}\left(\frac{n}{t}+1\right)-\ell(t^{n}e^{t})^{-\frac{3}{m}}
\end{equation}

It is observed that the spatial volume is zero and expansion scalar is infinite at 
$t = 0$, which shows that the universe starts evolving with zero volume at initial epoch at $t = 0$ with 
an infinite rate of expansion. The scale factor also vanish at initial moment hence the model 
has a point type singularity at $t = 0$. For $t \rightarrow \infty$, we get $q = -1$ and $\frac{dH}{dt} = 0$, 
which implies the greatest value Hubble's parameter and fastest rate of expansion of the universe. It is evident 
that negative value of $q$ would accelerate and increase the age of universe. 
Figure 1 shows the dynamics of DP versus cosmic time. It is observed that initially the DP evolves with positive 
sign but later on, DP grows with negative sign. This behaviour of DP clearly explain 
the decelerated expansion in past and accelerated expansion of universe at present as observed in 
recent observation of SN Ia. 
Thus the derived model can be utilized to describe the dynamics of the late time evolution of the observed universe.\\ 

In the derived model, the present value of DP is estimated as
\begin{equation}
\label{eq30}
q_{0}=-1+\frac{n}{mH_{0}^{2}t_{0}^{2}}
\end{equation}
Where $H_{0}$ is the present value of Hubble's parameter and $t_{0}$ is the age of universe at present epoch. If 
we set $n = 0.27m$ in equation (\ref{eq30}), we obtain $q_{0} = -0.73$ which is exactly match with the observed value 
of DP at present epoch (Cunha et al. 2009). Thus we constraint $m = 3$ and $n = 0.27m$ in the remaining 
discussions of the model and graphical representations of physical parameters.\\

Figure 2 depicts the variation of anisotropy parameter $(\bar{A})$ 
versus cosmic time. It is shown that $\bar{A}$ decreases with time and tends to zero for 
sufficiently large times. Thus the anisotropic behaviour of universe dies out on later times 
and the observed isotropy of universe can be achieved in derived model at present epoch.\\
\begin{figure}
\begin{center}
\includegraphics[width=3.5in]{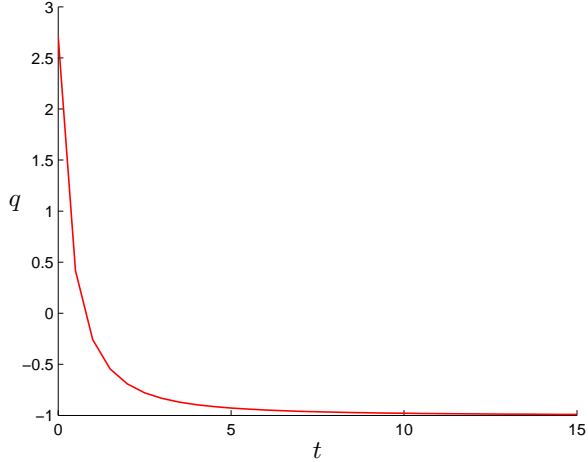} 
\caption{The plot of DP $(q)$ vs. time (t).}
\label{fg:asF1.eps}
\end{center}
\end{figure}

\begin{figure}
\begin{center}
\includegraphics[width=3.5in]{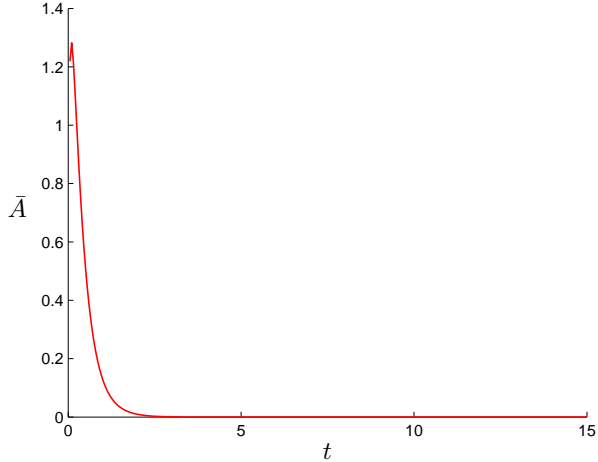} 
\caption{The plot of anisotropy parameter $(\bar{A})$ vs. time (t).}
\label{fg:as37F3.eps}
\end{center}
\end{figure}
  
The effective pressure $(\bar{p})$ and energy density $(\rho)$ of the model read as
\[
 \bar{p}=\alpha^{2}(t^{n}e^{t})^{-\frac{2}{m}}-2\left[\frac{1}{m}\left(\frac{n}{t}+1\right)+
\ell(t^{n}e^{t})^{-\frac{3}{m}}\right]^{2}-
\]
\begin{equation}
\label{eq31}
\left[\frac{1}{m^{2}}\left(\frac{n}{t}+1\right)^{2}-
\ell^{2}(t^{n}e^{t})^{-\frac{6}{m}}\right]+\Lambda
\end{equation}
\[
 \rho=\frac{3}{m^{2}}\left(\frac{n}{t}+1\right)^{2}-\ell^{2}(t^{n}e^{t})^{-\frac{6}{m}}-\;\;\;\;\;\;\;\;\;\;\;\;\;
\]
\begin{equation}
\label{eq32}
3\alpha^{2}(t^{n}e^{t})^{-\frac{2}{m}}-\Lambda
\end{equation}

Equations (\ref{eq8}) and (\ref{eq31}) lead to
\[
 p-3\xi H=\alpha^{2}(t^{n}e^{t})^{-\frac{2}{m}}-2\left[\frac{1}{m}\left(\frac{n}{t}+1\right)+
\ell(t^{n}e^{t})^{-\frac{3}{m}}\right]^{2}
\]
\begin{equation}
\label{eq33}
-\left[\frac{1}{m^{2}}\left(\frac{n}{t}+1\right)^{2}-
\ell^{2}(t^{n}e^{t})^{-\frac{6}{m}}\right]+\Lambda
\end{equation}  

For the specification of $\xi$, we assume that the fluid obey the equation of state of the 
form
\begin{equation}
\label{eq34}
p=\gamma \rho
\end{equation}
where $(0\leq \gamma \leq 1)$ is constant. Thus, we can solve the cosmological parameters by taking 
different physical assumption on $\xi(t)$.\\

\section{Model with constant coefficient of bulk viscosity}
we assume that\\
$$\xi(t) = \xi_{0} = contant $$

Now, Equation (\ref{eq33}), with use of equations (\ref{eq32}) and (\ref{eq34}) reduces to
\[
 \rho=\frac{3\xi_{0}}{m(1+\gamma)}\left(\frac{n}{t}+1\right)+\frac{2}{m^{2}(1+\gamma)}\left(\frac{n}{t}+1\right)^{2}
\]
\begin{equation}
\label{eq35}
-\frac{2}{(1+\gamma)}\left[\frac{1}{m}\left(\frac{n}{t}+1\right)+\ell(t^{n}e^{t})^{-\frac{3}{m}}\right]^{2}-
2\alpha^{2}(t^{n}e^{t})^{-\frac{2}{m}}
\end{equation}
Eliminating $\rho(t)$ between equations (\ref{eq32}) and (\ref{eq35}), we get
\[
 \Lambda=\frac{(3\gamma+1)}{m^{2}(\gamma+1)}\left(\frac{n}{t}+1\right)^{2}+
2\left[\frac{1}{m}\left(\frac{n}{t}+1\right)+\ell^{2}(t^{n}e^{t})^{-\frac{3}{m}}\right]^{2}
\]
\begin{equation}
\label{eq36}
-\frac{3\xi_{0}}{\gamma+1}\left(\frac{n}{t}+1\right)-\alpha^{2}(t^{n}e^{t})^{-\frac{2}{m}}-\ell^{2}
(t^{n}e^{t})^{-\frac{6}{m}}
\end{equation}
From equation (\ref{eq36}), we observe that the cosmological 
constant is decreasing function of time and it approaches a 
small positive value as time progresses (i. e. present epoch).
\section{Model with bulk viscosity proportional to the energy density i. e. $\xi\propto \rho$}
We assume that
\begin{equation}
\label{eq37}
\xi=\xi_{0}\rho
\end{equation}
Firstly Murphy (1973) has constructed a class of viscous cosmological models with 
$\xi = \xi_{0}\rho$ which possesses an interesting feature that the big bang type 
singularity of infinite space-time curvature does not occurs at finite past. Later on 
Pradhan et al (2004, 2005) presented bulk viscous cosmological models for $\xi = \xi_{0}\rho$ in 
harmony with SN Ia observations.\\
\begin{figure}
\begin{center}
\includegraphics[width=3.5in]{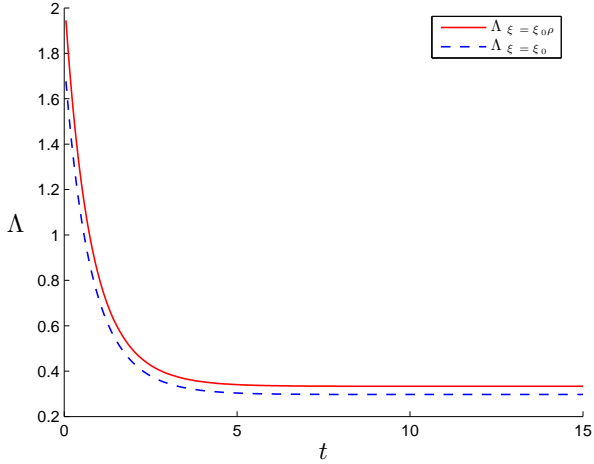} 
\caption{The plot of cosmological constant $(\Lambda)$ vs. time (t).}
\label{fg:as37F2.eps}
\end{center}
\end{figure}

Now, equation (\ref{eq33}) with use of equations (\ref{eq32}), (\ref{eq34}) and (\ref{eq37}) reduces to
\[
 \left[1+\gamma-\frac{3\xi_{0}}{m}\left(\frac{n}{t}+1\right)\right]\rho=\frac{2}{m^{2}}\left(\frac{n}{t}+1\right)^{2}-
\]
\begin{equation}
 \label{eq38}
2\left[\frac{1}{m}\left(\frac{n}{t}+1\right)+\ell(t^{n}e^{t})^{-\frac{3}{m}}\right]^{2}-
2\alpha^{2}(t^{n}e^{t})^{-\frac{2}{m}}
\end{equation}
Eliminating $\rho(t)$ between equations (\ref{eq32}) and (\ref{eq38}), we obtain
\[
 \Lambda=\frac{3}{m^{2}}\left(\frac{n}{t}+1\right)^{2}-\ell^{2}(t^{n}e^{t})^{-\frac{6}{m}}-
3\alpha^{2}(t^{n}e^{t})^{-\frac{2}{m}}
\]
\[
 -\frac{2}{m^{2}}\left(\frac{n}{t}+1\right)^{2}+\frac{1}{1+\gamma-\frac{3\xi_{0}}{m}\left(\frac{n}{t}+1\right)}\times
\]
\begin{equation}
\label{eq39}
\left[2\alpha^{2}(t^{n}e^{t})^{-\frac{2}{m}}
+\frac{2}{m}\left(\frac{n}{t}+1\right)+\ell(t^{n}e^{t})^{-\frac{3}{m}}\right]^{2}
\end{equation}

From equation (\ref{eq39}), it is observed that the cosmological term is positive and decreasing function 
of time (i. e. present epoch) which supports the result obtained from recent type Ia supernova observations 
(Riess et al. 1998; Perlmutter at al. 1997). The behaviour of cosmological constant is clearly shown in 
Figure 3. The models have non-vanishing cosmological constant and energy density as $t \rightarrow \infty$. 
It is well known that with the expansion of universe i. e. with the increase of time $t$, the 
energy density decreases and becomes too small to be ignored.\\  

We can express equations (\ref{eq10})$-$(\ref{eq13}) in terms of $H$, $q$ and $\sigma$ as
\begin{equation}
\label{eq40}
\bar{p}-\Lambda=(2q-1)H^{2}-\sigma^{2}+\frac{\alpha^{2}}{a^{2}}
\end{equation}
\begin{equation}
\label{eq41}
\rho+\Lambda=3H^{2}-\sigma^{2}-\frac{3\alpha^{2}}{a^{2}}
\end{equation}
From equations (\ref{eq40}) and (\ref{eq41}), we obtain
\begin{equation}
\label{eq42}
\frac{\ddot{a}}{a}=\frac{\Lambda}{3}+\frac{1}{2}\xi\theta-\frac{1}{6}(\rho+3p)-\frac{2}{3}\sigma^{2}
\end{equation}
which is Raychaudhuri's equation for given distribution. Equation (\ref{eq42}) shows that 
for $\rho+3p = 0$, acceleration is initiated by bulk viscosity and $\Lambda$ term. In absence of 
bulk viscosity only $\Lambda$ contributes the acceleration that seems to relate $\Lambda$ with dark energy. 
It also shows that for a positive $\Lambda$ the universe may accelerate with the condition $\rho+3p\leq 0$ i. e. 
p is negative for positive energy density $(\rho)$ with 
a definite contribution of $\Lambda$ in the acceleration. In the observational front, the data 
set coming from supernova legacy survey (SNLS) show that the dark energy behaves in the same manner as that 
$\Lambda$.  
\section{Concluding remarks}
In this paper, we have presented the generalized law for scale factor in homogeneous and anisotropic 
Bianchi-V space-time that yields the time dependent DP, representing a model which generates a transition 
of universe from early decelerating phase to recent accelerating phase. The spatial scale factors 
and volume scalar vanish at $t = 0$. The energy density and pressure are infinite at this initial epoch. 
As $t \rightarrow \infty$, the scale factor diverge and $\rho$, $p$ both tend to zero. $\bar{A}$ 
and $\sigma^{2}$ are very large at initial moment but decrease with cosmic time and vanish 
at $t \rightarrow \infty$. The model shows isotropic state in later time of its evolution. Also we observe 
that $\bar{p} = -\rho$ as $t \rightarrow\infty$. For $n\neq 0$, all matter and radiation is concentrated 
at the big bang epoch and the cosmic expansion is driven by the big bang impulse. The model has a point 
type singularity at the initial moment as the scale factors and volume vanish at $t = 0$. For $n = 0$, 
the model has no real singularity and energy density becomes finite. Thus the universe has 
non singular origin and the cosmic expansion is driven by the creation of matter particles. 
It has been observed that $lim_{t\rightarrow 0}\frac{\rho}{\theta^{2}}$ turns out to be constant. 
Thus the model approaches homogeneity and matter is dynamically negligible near the origin.\\   

The cosmological constant given in sections $4$ and $5$ are decreasing function 
of time and they are approach a small positive value as time increases (i. e. present epoch). 
The value of cosmological constant for these models are supported by the results from recent 
supernovae observations recently obtained by the High-Z Supernovae Team and Supernovae Cosmological 
Project (Perlmutter et al. 1997, 1998, 1999; Riess et al. 1998, 2004). A positive cosmological constant 
resists the attractive gravity of matter due to its negative pressure. For most universes, 
the positive cosmological constant eventually dominates over the attraction of matter and drives 
the universe to expands exponentially. Thus, with our approach, we obtain a physically relevant 
decay law for the cosmological constant unlike other authors where adhoc laws were used to arrive 
at a mathematical expressions for decaying $\Lambda$. Thus the derived models are more general than 
those studied earlier.\\ 


\end{document}